\documentclass[journal=jacsat,manuscript=article]{achemso}
\usepackage{chemformula} % Formula subscripts using \ch{}
\usepackage[T1]{fontenc} % Use modern font encodings

\usepackage{float}
\usepackage{graphicx}
\usepackage{pdfpages}
\usepackage{dcolumn}
\usepackage{bm}
\usepackage{mathrsfs,amssymb,amsmath}
\usepackage{ulem}
\usepackage{caption}
\usepackage[colorlinks=true,linkcolor=blue,citecolor=blue,anchorcolor=blue,urlcolor=blue]{hyperref}
\usepackage{booktabs}

\usepackage{setspace}
\usepackage{url}
\usepackage{pdfpages}
\author{Yan Yin}
\affiliation[NUAA]
{State Key Laboratory of Mechanics and Control for Aerospace Structures \& Key Laboratory for Intelligent Nano Materials and Devices of Ministry of Education \& Institute for Frontier Science \& College of Aerospace Engineering, Nanjing University of Aeronautics and Astronautics (NUAA), Nanjing 210016, China}

\author{Qihua Gong}
\email{gongqihua@nuaa.edu.cn}
\affiliation[NUAA]
{State Key Laboratory of Mechanics and Control for Aerospace Structures \& Key Laboratory for Intelligent Nano Materials and Devices of Ministry of Education \& Institute for Frontier Science \& College of Aerospace Engineering, Nanjing University of Aeronautics and Astronautics (NUAA), Nanjing 210016, China}
\alsoaffiliation[NUAA2]{MIIT Key Laboratory of Aerospace Information Materials and Physics \& College of Physics, Nanjing University of Aeronautics and Astronautics (NUAA), Nanjing 211106, China}

\author{Min Yi}
\email{yimin@nuaa.edu.cn}
\affiliation[NUAA]
{State Key Laboratory of Mechanics and Control for Aerospace Structures \& Key Laboratory for Intelligent Nano Materials and Devices of Ministry of Education \& Institute for Frontier Science \& College of Aerospace Engineering, Nanjing University of Aeronautics and Astronautics (NUAA), Nanjing 210016, China}

\author{Wanlin Guo}
\affiliation[NUAA]
{State Key Laboratory of Mechanics and Control for Aerospace Structures \& Key Laboratory for Intelligent Nano Materials and Devices of Ministry of Education \& Institute for Frontier Science \& College of Aerospace Engineering, Nanjing University of Aeronautics and Astronautics (NUAA), Nanjing 210016, China}

\date{\today}
\title[AnTitle]{Monolayer polar metals with large piezoelectricity derived from MoSi$_2$N$_4$}

\abbreviations{IR,NMR,UV}

\begin{document}    
%% Here goes the abstract

\begin{abstract}
The advancement of two-dimensional polar metals tends to be limited by the incompatibility between electric polarity and metallicity as well as dimension reduction.
Here, we report polar and metallic Janus monolayers of MoSi$_2$N$_4$ family by breaking the out-of-plane (OOP) structural symmetry through Z (P/As) substitution of N. Despite the semiconducting nature of MoSi$_2$X$_4$ (X=N/P/As), four Janus MoSi$_2$N$_{x}$Z$_{4-x}$ monolayers are found to be polar metals owing to the weak coupling between the conducting electrons and electric polarity. The metallicity is originated from the Z substitution induced delocalization of occupied electrons in Mo-d orbitals. The OOP electric polarizations around 10--203~pC/m are determined by the asymmetric OOP charge distribution due to the non-centrosymmetric Janus structure. The corresponding OOP piezoelectricity is further revealed as high as 39--153~pC/m and 0.10--0.31~pm/V for piezoelectric strain and stress coefficients, respectively. The results demonstrate polar metallicity and high OOP piezoelectricity in Janus MoSi$_2$N$_{x}$Z$_{4-x}$ monolayers and open new vistas for exploiting unusual coexisting properties in monolayers derived from MoSi$_2$N$_4$ family.
\end{abstract}

\\[1.5em]

% Main text
\section{1. INTRODUCTION}
Metals with electric polarization have emerged to exhibit fantastic physical properties, e.g., ferroelectricity, superconductivity and magnetoelectricity~\cite{Shi2013-NatMater, Puggioni2014-noncenstrosymmetric-NC, Kim2016-design-Nature, Benedek2016-reexamine-JMCC, Cao2018-ArtificialPolarMetal-NC, Laurita2019-NC, Meng2019-MagneticPolarMetal-NC, Cai2021-antiperovskites-NC}.
Polar metals are considered as a more appropriate way to define the metals in a polar space group before the polarization switchability remains unresolved~\cite{Ke2021-MaterialsHorizons}. Several two-dimensional (2D) polar metals have been reported, for instance, the tri-layer superlattices BaTiO$_3$/SrTiO$_3$/LaTiO$_3$~\cite{Cao2018-ArtificialPolarMetal-NC}, the multiferroic metal $\alpha$-In$_2$Se$_3$~\cite{Duan2021-In2Se3-MaterialsHorizons} and the artificial ferroelectric Ba$_{0.2}$Sr$_{0.8}$TiO$_3$ thin films~\cite{Zhou2019-BaSrTiO-CommunicationPhysics}. Cobden et al. found that non-polar topological semimetal WTe$_2$ exhibits spontaneous out-of-plane (OOP) electric polarization and metallicity through two- or three-layer stacking~\cite{Fei2018-WTe2-Nature}. 
Recently, a large family of 2D bimetal phosphates have been systematically investigated, including 16 ferroelectric metals~\cite{Ma2021-BimetalPhosphats-ScienceBulletin}. The ferroelectricity is attributed to the break of spontaneous symmetry induced by the opposite vertical displacements of different bimetal atoms. The previous studies have no doubt revealed the intriguing phenomena in 2D polar metals and boosted the development of other materials.
Currently, seeking more 2D polar metals still deserves more efforts and the pace has never stopped.

%\textbf{2. MoSi$_{2}$N$_{4}$ systems}
MoSi$_{2}$N$_{4}$ is a novel 2D material that has been experimentally synthesized~\cite{Hong2020-MoSi2N4-Science} and is further expanded to MA$_{2}$Z$_{4}$ family in theoretical investigations~\cite{Wang2021-MAZ-NatCommun}. In MA$_{2}$Z$_{4}$, M is Group \uppercase\expandafter{\romannumeral4}B -- \uppercase\expandafter{\romannumeral6}B transition-metal elements, A is Si or Ge and Z is Group \uppercase\expandafter{\romannumeral5}A elements (N/P/As).
The element selectivity brings structural diversity of MA$_{2}$Z$_{4}$ and further renders the family with intriguing chemical and physical properties which have been involved in the fields of electronics, optoelectronics, thermal transport and spintronics~\cite{Cao2021-electricalContact-APL, Yang2021-electronic-optical-Nanoscale, Yao2021-optoelectronic-nanomaterials, Huang2022-AdvOptMater, Zhong2021-SlidingFerroelectricity-JMCA, Yu2021-thermal-NewJPhys, Yin2021-thermal-ACSAMI, Wang2021-MAZ-NatCommun, Li2020-valley-PRB, Chen2021-Magnetic-Chemistry, Li2021-VSi2N4-AnnalenderPhysik, Akanda2021-VSi2N4-CrSi2N4-APL, Yin2022-arxiv-MAZreview}. 
It is found that semiconducting or metallic character of monolayer MA$_{2}$Z$_{4}$ is dependent on the valence electron counts (VECs) of the system~\cite{Wang2021-MAZ-NatCommun, Zhou2021-Symmetry-JPCL}, where the transition-metal atom rather than the structural phase plays a key role. For instance, the majority of MA$_{2}$Z$_{4}$ with Group \uppercase\expandafter{\romannumeral4}B or \uppercase\expandafter{\romannumeral6}B elements (32 or 34 VECs in total) are semiconductors, while those with \uppercase\expandafter{\romannumeral5}B elements (33 VECs in units cell) are metals. However, the influence of structural phase on MA$_{2}$Z$_{4}$ is rarely examined. 
Janus MSiGeN$_4$ (M~=~Mo/W) monolayers, are revealed as indirect bandgap semiconductors with a high electron mobility~\cite{Guo2021-MSiGeN4-JMCC}. Meanwhile, MSiGeN$_4$ monolayers have in-plane and OOP polarizations under a uniaxial strain and valley polarization under the spin-orbit coupling.
Recently, the transition from Stoner ferromagnetism to half-metal by hole doping in MoN$_2$X$_2$Y$_2$ (X, Y~=~Group \uppercase\expandafter{\romannumeral3}A or \uppercase\expandafter{\romannumeral6}A elements) is presented~\cite{Ding2022-MoN2X2Y2-ASS}.
Nevertheless, polar metallic members have so far never been reported in MA$_{2}$Z$_{4}$ family.

Motivated by the diversity and designability of MA$_{2}$Z$_{4}$ family, here we design 16 novel MoSi$_2$N$_{x}$Z$_{4-x}$ (N~=~P/As) monolayers through Z substitution of N in MoSi$_2$N$_4$ by first-principles calculations. 13 MoSi$_2$N$_x$Z$_{4-x}$ structures are found stable in terms of formation energy and lattice stability, among which there are 6 semiconductors and 7 metals. Particularly, although MoSi$_2$X$_4$ (X=N/P/As) monolayers are all semiconductors, 4 kinds of Janus MoSi$_2$N$_{x}$Z$_{4-x}$ monolayers with non-centrosymmetry along OOP direction are demonstrated to be polar metals.
The metallicity is attributed to the Z substitution induced delocalization of electrons around Mo atoms and thus the obvious honeycomb-like conducting network. 
The OOP polarization ($P_\text{out}$) of these metallic Janus monolayers are predicted as 10--203~pC/m, which are higher than that of most 2D materials. 
The excellent OOP piezoelectricity of these polar metallic Janus MoSi$_2$N$_x$Z$_{4-x}$ monolayers is also expounded, with the piezoelectric strain and stress coefficients around 39--153~pC/m and 0.10--0.31~pm/V, respectively. 
These results advocate Janus MoSi$_2$N$_x$Z$_{4-x}$ monolayers as polar metals with large piezoelectricity, providing new opportunities for realizing 2D multifunctional materials with unusual coexisting properties.

\begin{figure}[!t]
\centering
\includegraphics[width=12cm]{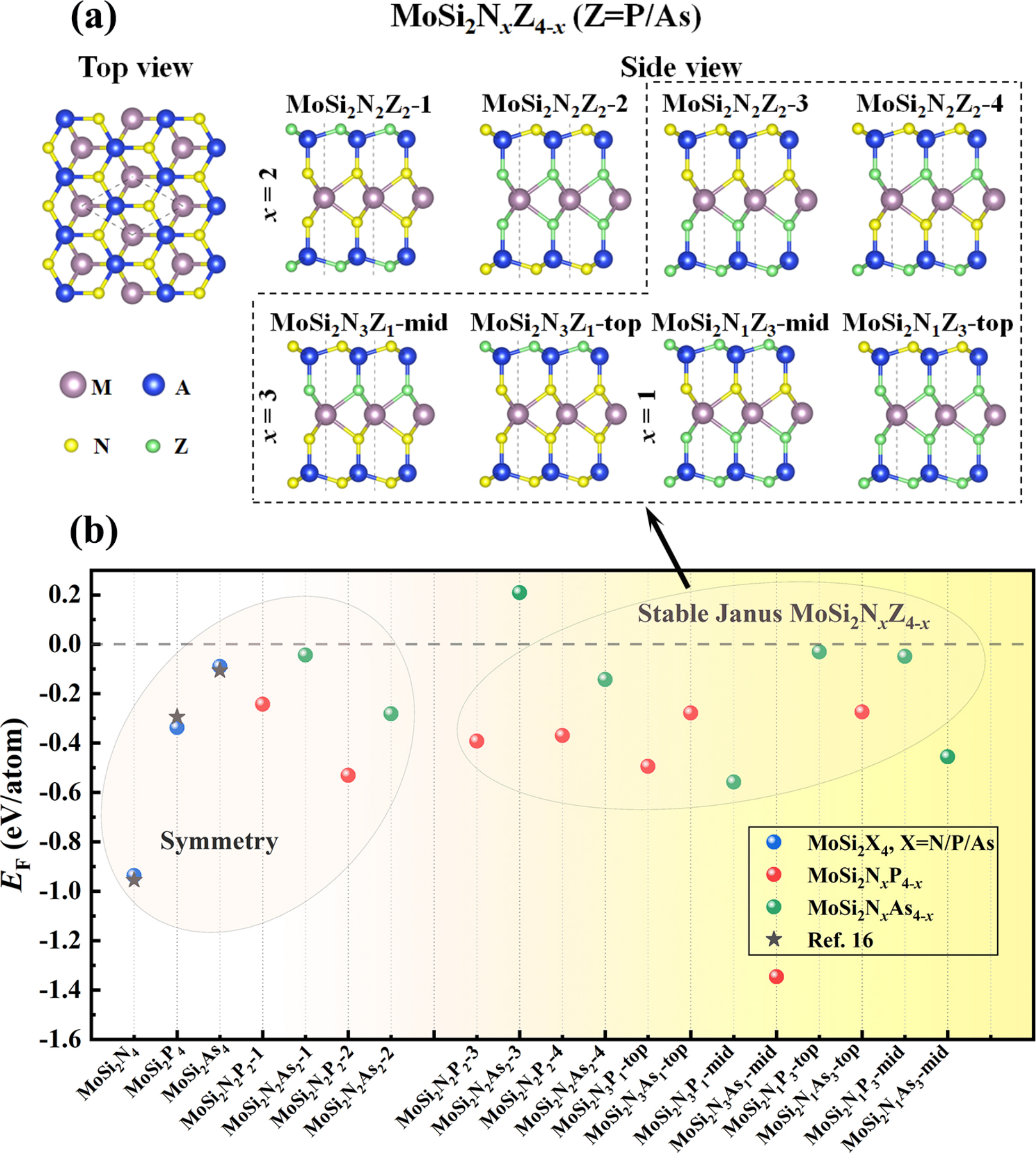}
\caption{(a) Top and side view of lattice structure. (c) Formation energy ($E_{\text{F}}$) of MoSi$_2$X$_4$ (X~=~N/P/As) and MoSi$_2$N$_x$Z$_{4-x}$ ($x$~=~1,~2,~3,~4) monolayers.}
\label{structure}
\end{figure}

%\textit{\textbf{Lattice structure and stability}}
%\textbf{Lattice structure and stability} -- 
\section{2. RESULTS AND DISCUSSION}
\subsection{2.1. Lattice structure and stability}
Monolayer MoSi$_2$N$_{4}$, a 7-atomic-layer structure with a hexagonal unit cell, is built by intercalating a 2H-MoS$_2$-type MoN$_2$ layer into an ${\alpha}$-InSe-type Si$_2$N$_2$. It can be also considered as a structure with mirror symmetry centred on Mo atom. Four N atomic layers locate at the two side of Mo layer, and Si atoms bridge N atoms from the side view. 
Here, after substituting N with P/As atoms, monolayer MoSi$_2$N$_{x}$Z$_{4-x}$ ($x$~=~1,~2,~3,~4) possesses the similar configuration, as shown in Fig.~\ref{structure}(a).
Four layers of Z (Z~=~P/As) atoms are alternated with Mo and Si layers, resulting in 18 kinds of lattice structures. The corresponding substitution ratio ($R_{s}$) for N is 0$\%$ (reference phase MoSi$_2$N$_4$), 25$\%$ (MoSi$_2$N$_3$Z$_1$-mid, MoSi$_2$N$_3$Z$_1$-top), 50$\%$ (MoSi$_2$N$_2$Z$_2$-1, MoSi$_2$N$_2$Z$_2$-2, MoSi$_2$N$_2$Z$_2$-3, MoSi$_2$N$_2$Z$_2$-4), 75$\%$ (MoSi$_2$N$_3$Z$_1$-mid, MoSi$_2$N$_3$Z$_1$-top) and 100$\%$ (previously reported MoSi$_2$P$_4$ and MoSi$_2$As$_4$~\cite{Wang2021-MAZ-NatCommun}). 
When $R_{s}$~=~0 or 100$\%$, the lattice belongs to the space group of $P\overline{6}m2$ (No.~187) and shows ``high-symmetry phase''. With Z atoms substitution, four types (MoSi$_2$N$_2$P$_2$-1, MoSi$_2$N$_2$P$_2$-2, MoSi$_2$N$_2$As$_2$-1, and MoSi$_2$N$_2$As$_2$-2) remain this mirror symmetry. Since the reduction of spontaneous geometric symmetry, other 12 structures with different Z atoms are subordinate to $P3m1$ (No.~156) group.
%For conveniently distinguishing, we classify them as two groups on the basis of mirror symmetry or not, which are symmetry (7 kinds) and asymmetry (12 kinds). 
The determined lattice parameters are listed in Table~S1. One can find that in pure MoSi$_2$X$_4$ the lattice constants, effective thickness ($h$) and bond lengths increase with the increasing period number of elements Z from N, P to As. The similar phenomenon also occurs in MoSi$_2$N$_x$Z$_{4-x}$ and is more notable when Z (P/As) locates at the outermost side.

\begin{figure*}[!t]
\centering
\includegraphics[width=14cm]{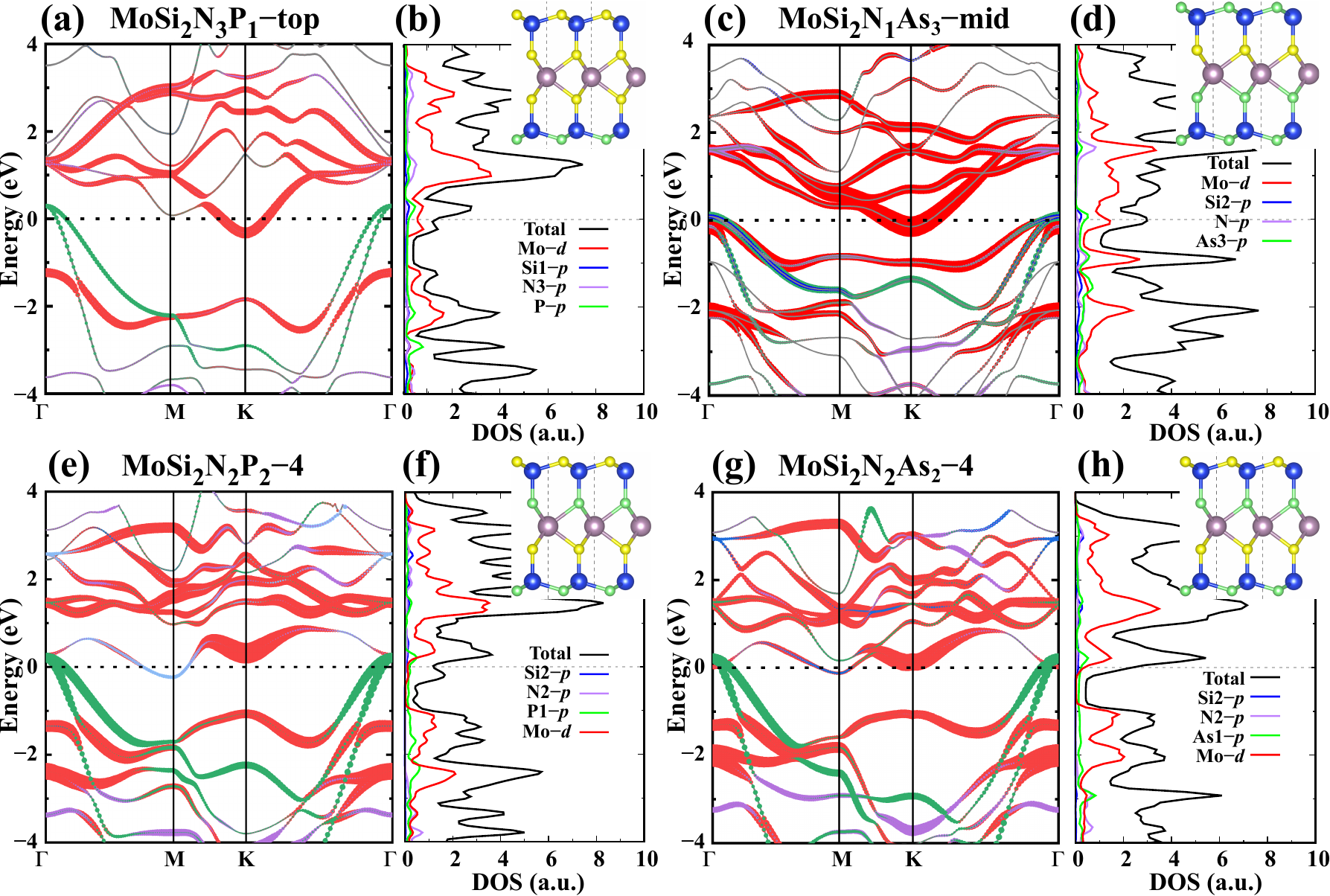}
\caption{Band structure and DOS of (a) (b) MoSi$_2$N$_3$P$_1$-top, (c) (d) MoSi$_2$N$_1$As$_3$-mid, (e) (f) MoSi$_2$N$_2$P$_2$-4, (g) (h) MoSi$_2$N$_2$As$_2$-4. The dot lines at zero energy represent the Fermi level.}
\label{band-all}
\end{figure*}

Then, we explore the stability of monolayer MoSi$_2$N$_x$Z$_{4-x}$. %The formation energy ($E_\text{F}$) of MoSi$_2$N$_x$Z$_{4-x}$ is expressed as 
%\begin{equation}
%E_{\text{F}}=\frac{E_{\text{MoSi$_2$N$_x$Z$_{4-x}$}}-(E_{\text{Mo}}+2E_{\text{Si}}+xE_{\text{N}}+(4-x)E_{\text{Z}})} {N_\text{atom}} 
%\end{equation}
%\noindent
%where $E_{\text{Mo}}$, $E_{\text{Si}}$, $E_{\text{N}}$ and $E_{\text{Z}}$ are energy of isolated Mo, Si, N and Z atom, respectively. $E_{\text{MoSi$_2$N$_x$Z$_{4-x}$}}$ is the energy of MoSi$_2$N$_x$Z$_{4-x}$. $N_\text{atom}$ is the number of total atoms.
%The enthalpies of formation ($E_\text{F}$) 
%The calculated $E_\text{F}$ is summarized in Fig.~\ref{structure}(b) and Table~S1.
The formation energy ($E_\text{F}$) of MoSi$_2$N$_x$Z$_{4-x}$ is summarized in Fig.~\ref{structure}(b) and Table~S1.
$E_\text{F}$ of MoSi$_2$N$_4$, MoSi$_2$P$_4$ and MoSi$_2$As$_4$ are -0.94, -0.34 and -0.09~eV/atom, respectively, agreeing well with the previous reports~\cite{Wang2021-MAZ-NatCommun}. The negative $E_\text{F}$ demonstrates the exothermic reaction and the possibility of experimental existence. Thus, 15 MoSi$_2$N$_{x}$Z$_{4-x}$ monolayers with negative $E_\text{F}$ are energetically stable and only one (MoSi$_2$N$_{2}$As$_{2}$-3) with positive $E_\text{F}$ is unstable. On the other hand, the phonon dispersion spectra of MoSi$_2$N$_{x}$Z$_{4-x}$ (Fig.~S1 and S2) show that 13 MoSi$_2$N$_{x}$Z$_{4-x}$ and 3 MoSi$_2$X$_{4}$ (X=N/P/As) monolayers without imaginary frequency are dynamically stable. Therefore, in addition to MoSi$_2$X$_{4}$, MoSi$_2$N$_{x}$Z$_{4-x}$ monolayers are predicted as new and stable counterparts that are derived from MoSi$_2$N$_4$ in terms of formation energy and lattice dynamics. Meanwhile, there are 9 kinds of Janus MoSi$_2$N$_{x}$Z$_{4-x}$ monolayers as shown in Fig.~\ref{structure}(b) and Table S1.

\subsection{2.2 Metallic Janus MoSi$_2$N$_{x}$Z$_{4-x}$ monolayers}

%\textit{\textbf{Metallic Janus MoSi$_2$N$_{x}$Z$_{4-x}$ monolayers}} 
%\textbf{Metallic Janus MoSi$_2$N$_{x}$Z$_{4-x}$ monolayers} -- 
The electric properties of Janus MoSi$_2$N$_{x}$Z$_{4-x}$ monolayers are firstly investigated via band structure (Figs.~S4). 
Three MoSi$_2$X$_{4}$ semiconductors exhibit decreasing band gap from N, P to As (Fig.~S3), which are in good agreement with previous calculations~\cite{Wang2021-MAZ-NatCommun, Yin2021-thermal-ACSAMI}.
This manifests that the band gap of MoSi$_2$Z$_{4}$ reduces with the increasing atomic period in the same group. 
For other 13 stable MoSi$_2$N$_{x}$Z$_{4-x}$ monolayers, 6 are semiconductors with the band gap of 0.01--0.54~eV for PBE and 0.40--0.96~eV for HSE06 functional (Table~S1 and Fig.~S4). However, the band gap has no dependence on the lattice constant of MoSi$_2$N$_{x}$Z$_{4-x}$, which is different from MoSi$_2$X$_{4}$ and A-site-substituted MoN$_{2}$X$_2$Y$_2$~\cite{Ding2022-MoN2X2Y2-ASS}. Additionally, there exist 7 metals in MoSi$_2$N$_{x}$Z$_{4-x}$ monolayer, i.e., MoSi$_2$N$_{3}$P$_{1}$-top, MoSi$_2$N$_{2}$P$_{2}$-1, MoSi$_2$N$_{2}$P$_{2}$-4, MoSi$_2$N$_{2}$As$_{2}$-1, MoSi$_2$N$_{2}$As$_{2}$-2, MoSi$_2$N$_{2}$As$_{2}$-4 and MoSi$_2$N$_{1}$As$_{3}$-mid.  Among them, there are 4 metals with Janus structure, as shown in Fig.~\ref{band-all}.

\begin{figure}
\centering
\includegraphics[width=12cm]{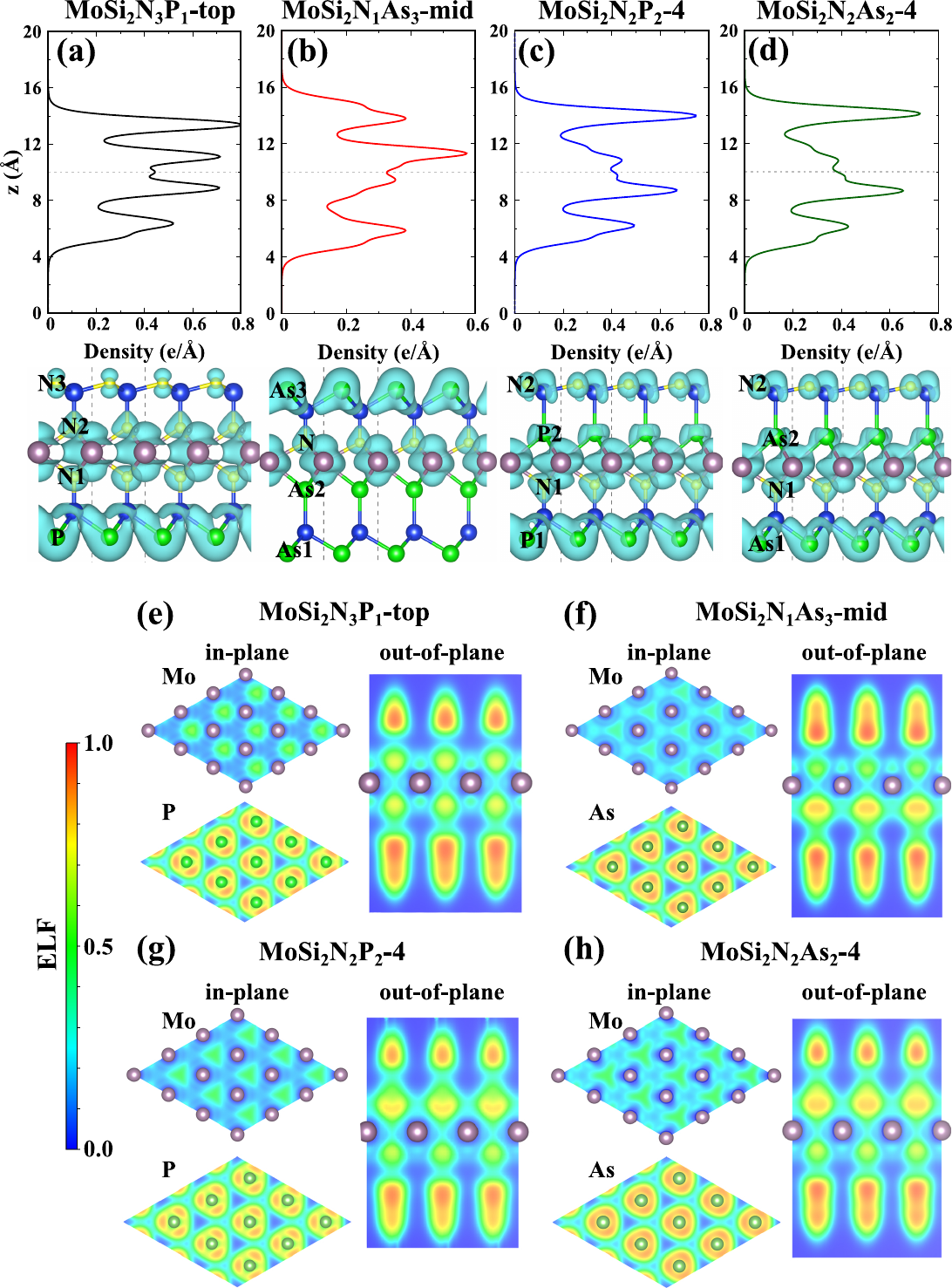}
\caption{Planar-average charge density along $z$-direction and partial charge density of the 16th, 17th, and 18th bands in the real place of four metal Janus monolayers: (a) MoSi$_2$N$_3$P$_1$-top, (b) MoSi$_2$N$_1$As$_3$-mid, (c) MoSi$_2$N$_2$P$_2$-4, (d) MoSi$_2$N$_2$As$_2$-4. The isovalue for charge density is 0.02 e/\AA$^{3}$. (e)--(h) Electron localization function (ELF) of in-plane (100) slice and out-of-plane (001) slice.}
\label{PartialCD-ProjectCD}
\end{figure}

To reveal the origin of metallic properties in MoSi$_2$N$_{x}$Z$_{4-x}$, we examine the projected band structure and density of states (DOS) of four Janus monolayers (Fig.~\ref{band-all}), i.e., MoSi$_2$N$_{3}$P$_{1}$-top, MoSi$_2$N$_{2}$P$_{2}$-4, MoSi$_2$N$_{2}$As$_{2}$-4, and MoSi$_2$N$_{1}$As$_{3}$-mid, which contains various substitution ratios and different OOP non-centrosymmetric structures.
Firstly, these four monolayers are all metals with three energy bands crossing Fermi level, which are the 16th, 17th and 18th band. From the partial DOS, it is found that these energy bands near the Fermi level are mainly occupied by electrons from Mo-$d$ orbitals in MoSi$_2$N$_{x}$Z$_{4-x}$, especially near the high-symmetry point of K. There also exists a little contribution from the $p$-orbital electrons of outermost P/As atoms on the side where N bridges Si and Mo. 
Secondly, MoSi$_2$N$_{2}$P$_{2}$-4 and MoSi$_2$N$_{2}$As$_{2}$-4 exhibit the same structural configuration (Z-Si-N-Mo-Z-Si-N) and thus the similar band structures, as shown in Fig.~\ref{band-all}~(e)--(h)). Actually, other MoSi$_2$N$_{x}$Z$_{4-x}$ monolayers with the same structural configuration show the similar energy bands as well (Fig.~S4). The electronic properties of different MoSi$_2$N$_{x}$Z$_{4-x}$ monolayers, however, are different. For instance, MoSi$_2$N$_{2}$P$_{2}$-2 and MoSi$_2$N$_{2}$P$_{2}$-3 are semiconductors, whereas MoSi$_2$N$_{2}$As$_{2}$-2 and MoSi$_2$N$_{2}$As$_{2}$-3 are metals. 
The similar energy bands reflect not only the similar effect of Z atoms on electronic properties in MoSi$_2$N$_x$Z$_{4-x}$, but also the ability of P/As atoms to reduce the band gap. 
Thirdly, in centrosymmetric MoSi$_2$N$_{4}$, the middle plane where Mo atom locates perpendicularly bisects the structure and results in the symmetric distribution of charges. However, the four metallic Janus structures are non-centrosymmetric, implying that there would exist intriguing phenomena caused by the OOP asymmetric charge distribution.

In order to further anatomize the electronic properties of the four Janus metals, we calculate the planar-average charge density along the OOP direction and the charge distribution of three energy bands crossing Fermi level in the real space, as shown in Fig.~\ref{PartialCD-ProjectCD}(a)--(d). The planar-average charge density distribution is found to be asymmetric with respect to the (001) plane where Mo atoms are located.
From the partial charge density of 16th, 17th, 18th energy bands in Fig.~\ref{PartialCD-ProjectCD}(a)--(d), it can be found that the charge distribution are mainly around Mo atoms and the outermost P/As atoms on the side where N bridges Si and Mo, in accordance with the band structure and DOS results in Fig.~\ref{band-all}. 
We further examine the electron localization function (ELF) that could qualitatively describe the strength of electron localization, as shown in Fig.~\ref{PartialCD-ProjectCD}(e)--(h).
In all these four Janus monolayers, the in-plane ELF crossing Mo atom is
about 0.5, indicating the electrons around Mo atom are strongly delocalized. In contrast, ELF crossing P/As atoms is close to 1.0 and thus the electrons around P/As atoms are localized. The electron localization of outermost P/As atoms is related to the strong covalent-like Si-P and Si-As bond. 
Therefore, the metallicity of these four Janus monolayers could be ascribed to the delocalization nature of electrons around Mo atom~\cite{Zhang2013-3DBN}. Specially, in MoSi$_2$N$_1$As$_3$-mid, the delocalized electrons even connect together to form an honeycomb-like conducting network.

\subsection{2.3 Polar metal and piezoelectricity}
\begin{figure*}[!t]
\centering
\includegraphics[width=14cm]{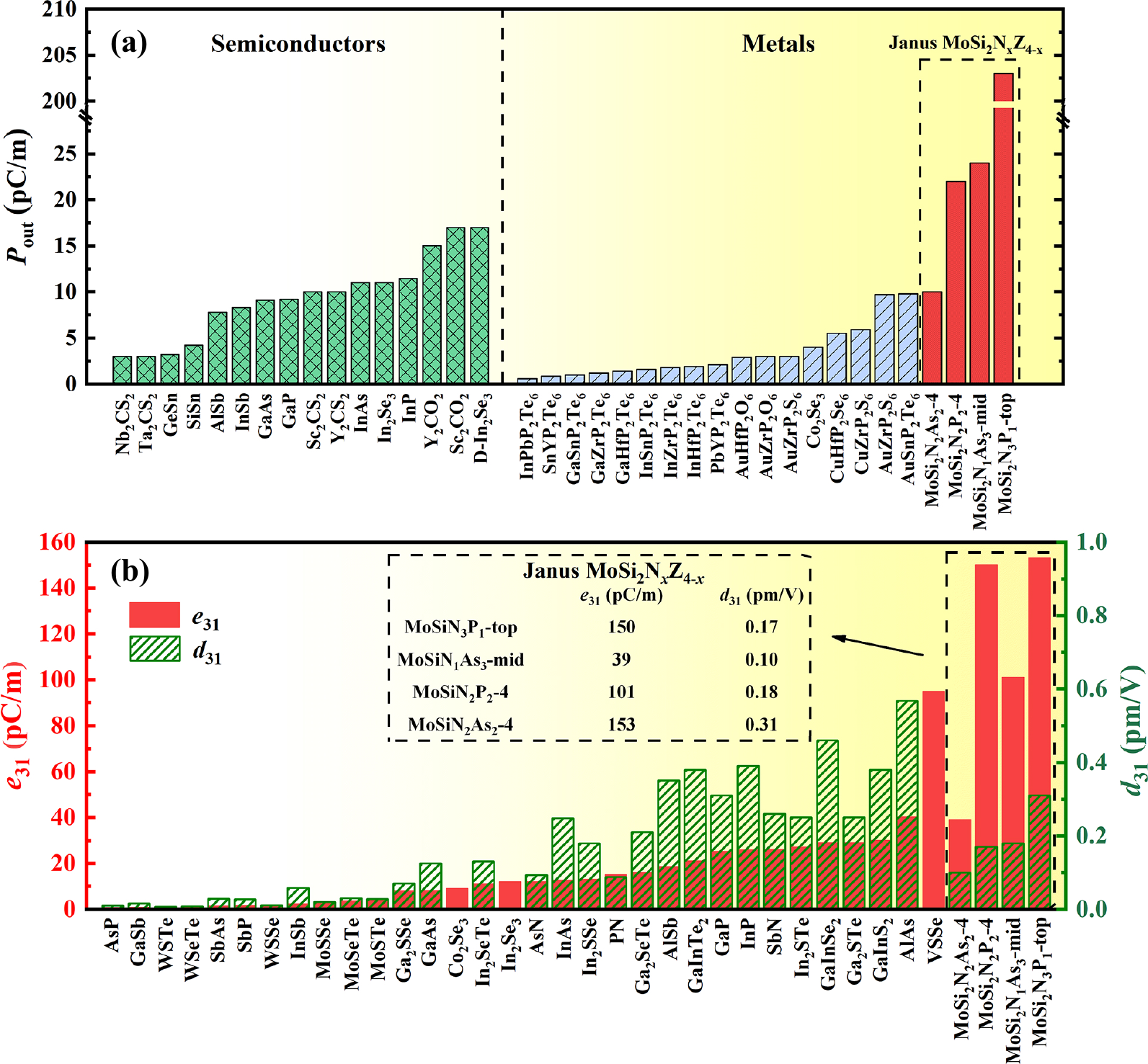}
\caption{Comparison of (a) OOP polarization ($P_\text{out}$), (b) piezoelectric strain ($e_{31}$) and stress ($d_{31}$) coefficients between  Janus metallic MoSi$_2$N$_x$Z$_{4-x}$ monolayers and other 2D materials~\cite{DiSante2015-PRB-BinaryCompounds, TangC2020-Nanoscale-First-principles, TangC2021-JPCC-ControllablePolarization, Ma2021-BimetalPhosphats-ScienceBulletin, Tang2021-JMCC-Co2Se3, Yagmurcukardes2019-PRB-JanusMoSTe, Zhang2019-NanoLett-VSSe, Chen2021-PRB-vdWHs-piezo}.}
%{ piezoelectric coefficients~\cite{Guo2017-APL-JanusGroupIII, Yin2017-JPCC-groupV-giant, Blonsky2015-groupIII-V-piezo, Dong2017-ACSNano-JanusTMDs, Tang2021-JMCC-Co2Se3, TangC2020-Nanoscale-First-principles, Zhang2019-NanoLett-VSSe, Chen2021-PRB-vdWHs-piezo}} 
%$P_{out}$ of MoSi$_2$N$_1$As$_3$-mid, MoSi$_2$N$_2$P$_2$-4 and MoSi$_2$N$_2$As$_2$-4 are close to other materials, but it of MoSi$_2$N$_3$P$_1$-top is more than ten folds higher than others.}
\label{2D-polarization}
\end{figure*}

%\textit{\textbf{Polar metal and piezoelectricity}}
In Table~\ref{table-piezo}, the absolute values of $P_\text{out}$ of Janus monolayer MoSi$_2$N$_3$P$_1$-top, MoSi$_2$N$_1$As$_3$-mid, MoSi$_2$N$_2$P$_2$-4, and MoSi$_2$N$_2$As$_2$-4 are 203, 22, 24, and 10~pC/m, respectively, which are more than ten folds higher than that of many 2D monolayer semiconductors (e.g., MXenes with $P_\text{out}$~=~(3--17)~pC/m~\cite{TangC2020-Nanoscale-First-principles}, group \uppercase\expandafter{\romannumeral4}A 2D binary ferroelectric compounds with $P_\text{out}$~=~(1--4)~pC/m~\cite{DiSante2015-PRB-BinaryCompounds}, and defective In$_2$Se$_3$ with  $P_\text{out}$~=~(1--4)~pC/m~\cite{TangC2021-JPCC-ControllablePolarization}) and 2D metals (e.g., 2D bimetal phosphates with $P_\text{out}$~=~(1--10)~pC/m~\cite{Ma2021-BimetalPhosphats-ScienceBulletin} and Co$_2$Se$_3$ with 4~pC/m~\cite{Tang2021-JMCC-Co2Se3}), as summarized in Fig.~\ref{2D-polarization}(a).  
The OOP polarization direction (insert figures in Fig.~S5) of four Janus metallic MoSi$_2$N$_x$Z$_{4-x}$ monolayers points to the out-of-plane direction, to the outermost P/As (on the side where N atom bridges Si and Mo). 
Such high $P_\text{out}$ provides these Janus MoSi$_2$N$_x$Z$_{4-x}$ metals a very competitive property among other 2D materials.

\begin{table}[htbp]
\centering
\caption{OOP dipole moment ($P_\text{out}$), elastic coefficient ($C_{11}$ and $C_{12}$) and OOP piezoelectric coefficients ($e_{31}$ and $d_{31}$) of four polar metallic Janus MoSi$_2$N$_x$Z$_{4-x}$ monolayers.}
\renewcommand\arraystretch{1.5} 
\setlength{\tabcolsep}{1.5mm}
\begin{tabular}{cccccc}%l=left, r=right,c=center
%\toprule
\hline
Structure &$P_\text{out}$ &$e_{31}$ &$C_{11}$ &$C_{12}$  &$d_{31}$ \\ %&$P_\text{out-berry}$ 
 &\multicolumn{2}{c}{(pC/m)} &(GPa) &(GPa) &(pm/V) \\ %&$\times$10$^{-10}$~C/m
\hline
MoSi$_2$N$_4$ &* &* &537.18 &159.70 &* \\ %&446.66 &1.18 \\

MoSi$_2$P$_4$ &* &* &171.16 &61.93 &* \\ %&773.46 &5.38 \\

MoSi$_2$As$_4$ &* &* &126.90 &42.74 &* \\ %&795.44 &6.78 \\

MoSi$_2$N$_3$P$_1$-top &203 &150 &313.29 &128.98 &0.17 \\

MoSi$_2$N$_1$As$_3$-mid	&22 &39 &141.84 &50.07 &0.10 \\

MoSi$_2$N$_2$P$_2$-4 &24 &101 &176.97 &104.79 &0.18 \\

MoSi$_2$N$_2$As$_2$-4 &10 &153 &169.42 &79.46 &0.31 \\
\hline
%\bottomrule
\end{tabular}
\label{table-piezo}
\end{table}

It is found that the piezoelectric effect can be generated by the electric polarization as well. The piezoelectricity is quantified by the coupling between polarization ($P_i$) and strain tensor ($\epsilon_{jk}$), i.e., $e_{ijk}$~=~$\partial{P_i}$/$\partial{\epsilon_{jk}}$~\cite{Wu2014-Nature-MoS2-piezo, Zhang2019-NanoLett-VSSe}. The subscript $i$, $j$ and $k$ denote the $x$, $y$ and $z$ direction. Here, we focus on the zigzag uniaxial strain effect on $P_\text{out}$, and investigate the OOP piezoelectric effect of these four Janus MoSi$_2$N$_x$Z$_{4-x}$ metals (Fig.~S5). The linear fitting of $P_\text{out}$ $vs.$ uniaxial strain data determines the OOP piezoelectric strain coefficients ($e_{31}$) as 150~pC/m for MoSi$_2$N$_3$P$_1$-top, 39~pC/m for MoSi$_2$N$_1$As$_3$-mid, 101~pC/m of MoSi$_2$N$_2$P$_2$-4, and 153~pC/m for MoSi$_2$N$_2$As$_2$-4. These $e_{31}$ values are quite larger than that of 1H-MoSTe (50~pC/m~\cite{Yagmurcukardes2019-PRB-JanusMoSTe}, half-metallic Co$_2$Se$_3$ (9~pC/m)~\cite{Tang2021-JMCC-Co2Se3} and Janus VSSe (98~pC/m)~\cite{Zhang2019-NanoLett-VSSe}, and are comparable to $e_{31}$ of InSe/TMDs van der Waals heterostructures (17--290~pC/m)~\cite{Chen2021-PRB-vdWHs-piezo}. 

To further investigate the piezoelectric stress coefficients ($d_{31}$), it is necessary to get the deformation  ability of the structures.
2D MoSi$_2$N$_4$ exhibits excellent mechanical properties as reported by previous experimental and theoretical studies~\cite{Hong2020-MoSi2N4-Science, Li2021-PE-MoSi2N4-mechanism}. The elastic coefficients of MoSi$_2$X$_4$ decrease from X~=~N to X~=~As (Table~\ref{table-piezo}), indicating N atoms play a crucial role in the rigidness of MA$_2$Z$_4$ family. When N atoms are replaced by P/As atoms, the rigidness of four MoSi$_2$N$_x$Z$_{4-x}$ metals is weaker than MoSi$_2$N$_4$, implying that they are easily deformed. $d_{31}$ of these four metals (with 3$m$ point-group symmetry) can be defined as $d_{31}$~=~$e_{31}$/($C_{11}$+$C_{12}$), where $C_{11}$ and $C_{12}$ are elastic constants~\cite{Fei2015-APL-piezo-cal, Yin2017-JPCC-groupV-giant}.
For MoSi$_2$N$_3$P$_1$-top, MoSi$_2$N$_1$As$_3$-mid, MoSi$_2$N$_2$P$_2$-4 and MoSi$_2$N$_2$As$_2$-4, $d_{31}$ are 0.17, 0.10, 0.18 and 0.31~pV/m, respectively. 
This OOP piezoelectricity $d_{31}$ is comparable to the highest $d_{31}$ that is reported so far for other 2D materials, as shown in Fig.~\ref{2D-polarization}(b).

\section{3. CONCLUSIONS}
In summary, we have reported polar and metallic Janus MoSi$_2$N$_x$Z$_{4-x}$ (Z=P/As) monolayers that are derived from MoSi$_2$N$_4$ by breaking the OOP structural symmetry through Z substitution of N. Among the 13 stable MoSi$_2$N$_x$Z$_{4-x}$ monolayers, there exist 6 semiconductors and 7 metals with four non-centrosymmetric Janus structures as polar metals. Although MoSi$_2$Z$_4$ (X=N/P/As) monolayers are semiconductors, these four Janus MoSi$_2$N$_x$Z$_{4-x}$ monolayers are found metallic.
The metallic nature is originated from the 16th, 17th and 18th energy bands crossing Fermi level, which are mainly occupied by electrons from Mo-$d$ orbitals. Analyzing charge density distribution and ELF reveal the conducting mechanism attributed to the delocalization of electrons around Mo atom. 
The non-centrosymmetric structure induces the asymmetric charge distribution along the OOP direction and thus the OOP electric polarization and piezoelectricity in four Janus metallic MoSi$_2$N$_x$Z$_{4-x}$ monolayers.
$P_\text{out}$, $e_{31}$, and $d_{31}$ are found up to 10--203~pC/m, 39--153~pC/m, and 0.10–0.31 pV/m, respectively, which are comparable to or even higher than the highest values that are reported so far for other 2D materials.
These results demonstrate MoSi$_2$N$_4$ derived Janus monolayers as polar metals with large OOP piezoelectricity, enrich the 2D polar metal family with coexisting unusual properties, and could inspire the asymmetric structure design of polar metals with large piezoelectricity. 

\section{4. COMPUTATIONAL METHODS}
First-principles calculations are performed by using Vienna $ab~initio$ Simulation Package (VASP) based on density functional theory (DFT)~\cite{Hafner2007-vasp, HAFNER2008-vasp}. The electron-ion interactions and exchange-correlation function are described by the projector augmented wave (PAW) method~\cite{Blochl1994-PAW} and the generalized gradient approximation (GGA) with Perdew–Burke–Ernzerhof (PBE)~\cite{Perdew1996-GGA}, respectively. For the confirmation of electric properties, e.g., band structures and semiconductor or metallic nature, the hybrid functional (HSE06)~\cite{Vydrov2006-HSE, Paier2006-HSE} is utilized. The cutoff energy of plane wave is set as 500~eV. The convergence criteria for energy and force are 10$^{-7}$~eV and 10$^{-4}$~eV/\AA, respectively. Monkhorst-Park $k$-mesh of a 21$\times$21$\times$1 for unit cell in the first Brillouin zone is used. The vacuum region is set as 20~\AA~to eliminate the interactions between layers. In MoSi$_2$X$_4$ (X=N/P/As) and MoSi$_2$N$_{x}$Z$_{4-x}$ monolayers, the valence electron configurations are treated as 5$s^2$4$d^4$ for Mo, 3$s^2$3$p^2$ for Si, 2$s^2$2$p^3$ for N, 3$s^2$3$p^3$ for P and 4$s^2$4$p^3$ for As. The structural stability is confirmed in terms of formation energy and lattice dynamics. 
The phonon dispersion spectra are obtained by the PHONOPY package with the finite displacement method~\cite{Togo2015-PHONOPY}. The 5$\times$5$\times$1 supercells and 5$\times$5$\times$1 $k$-meshes are used the calculation of harmonic interatomic force constants (2$^{nd}$ IFCs).

The formation energy ($E_\text{F}$) of MoSi$_2$N$_x$Z$_{4-x}$ is expressed as 
\begin{equation}
E_{\text{F}}=\frac{E_{\text{MoSi$_2$N$_x$Z$_{4-x}$}}-(E_{\text{Mo}}+2E_{\text{Si}}+xE_{\text{N}}+(4-x)E_{\text{Z}})} {N_\text{atom}} 
\end{equation}
\noindent
where $E_{\text{Mo}}$, $E_{\text{Si}}$, $E_{\text{N}}$ and $E_{\text{Z}}$ are energy of isolated Mo, Si, N and Z atom, respectively. $E_{\text{MoSi$_2$N$_x$Z$_{4-x}$}}$ is the energy of MoSi$_2$N$_x$Z$_{4-x}$.
$N_\text{atom}$ is the number of total atoms.

Since the Berry phase method is problematic for calculating polarizations of metals~\cite{Resta2002-InsulatorsMetals-JPCM, Xu2020-berryphaseErrorForMetals-NanoscaleHorizons}, we utilize an alternative method based on classical electrodynamics. 
In the case that the vacuum slab is large enough to eliminate the interlayer interactions and neglect the periodicity, the out-of-plane polarization ($P_\text{out}$) can be easily defined by the classical electrodynamics and estimated as $P~=~q~\times~d$, where $q$ is the total number of valence charge and $d$ is the vector of dipole moment from negative center (NCC) to positive center (PCC)~\cite{TangC2020-Nanoscale-First-principles, TangC2021-JPCC-ControllablePolarization, Tang2021-JMCC-Co2Se3}. The $z$ coordinates of NCC and PCC can be calculated as: 
\begin{equation}
\text{NCC}=\frac{\int{\int{\int{{\rho}zdxdydz}}}} {q}
\end{equation}
\noindent
\begin{equation}
\text{PCC}=\frac{\sum{nz}} {q}
\end{equation}
\noindent
in which $\rho$, $z$ and $n$ are the charge density, coordinate along the OOP $z$ direction and valence electrons of each ion, respectively. The calculated $P_\text{out}$ of monolayer $\alpha$-In$_2$Se$_3$ is 0.15~e\AA/u.c (18~pC/m), agreeing well with the previous measurements and theoretical calculations (0.095--0.18~e\AA/u.c)~\cite{TangC2020-Nanoscale-First-principles, TangC2021-JPCC-ControllablePolarization, Tang2021-JMCC-Co2Se3, Ding2017-NC-Predictionofintrinsic, Chen2021-PRB-vdWHs-piezo}. 
$P_\text{out}$ of CrSe$_2$-1H is calculated as 0.45~e\AA/u.c., in agreement with the calculation result (0.37~e\AA/u.c) as well~\cite{ZhangQF2022-JPCC-ModulatedFerromagnetism}. These testing calculations validate the applicability of this method.

\section*{Supporting information}
%\textit{\textbf{Supplementary}}
Supplementary material associated with this article can be found in the online version.

\section*{Acknowledgment}
%\textit{\textbf{Acknowledgment}} 
The authors acknowledge the support from the National Natural Science Foundation of China (NSFC 12272173, 11902150), 15th Thousand Youth Talents Program of China, the Research Fund of State Key Laboratory of Mechanics and Control for Aerospace Structures (MCMS-I-0419G01 and MCMS-I-0421K01), the Fundamental Research Funds for the Central Universities (1001-XAC21021), and a project Funded by the Priority Academic Program Development of Jiangsu Higher Education Institutions. This work is partially supported by High Performance Computing Platform of Nanjing University of Aeronautics and Astronautics. Simulations were also performed on Hefei advanced computing center.

%\section*{References}
%\bibliographystyle{apsrmp4-1custom}
%\bibliographystyle{unsrt}
\bibliography{References.bib}

\includepdf[pages=-,pagecommand={},width=18cm]{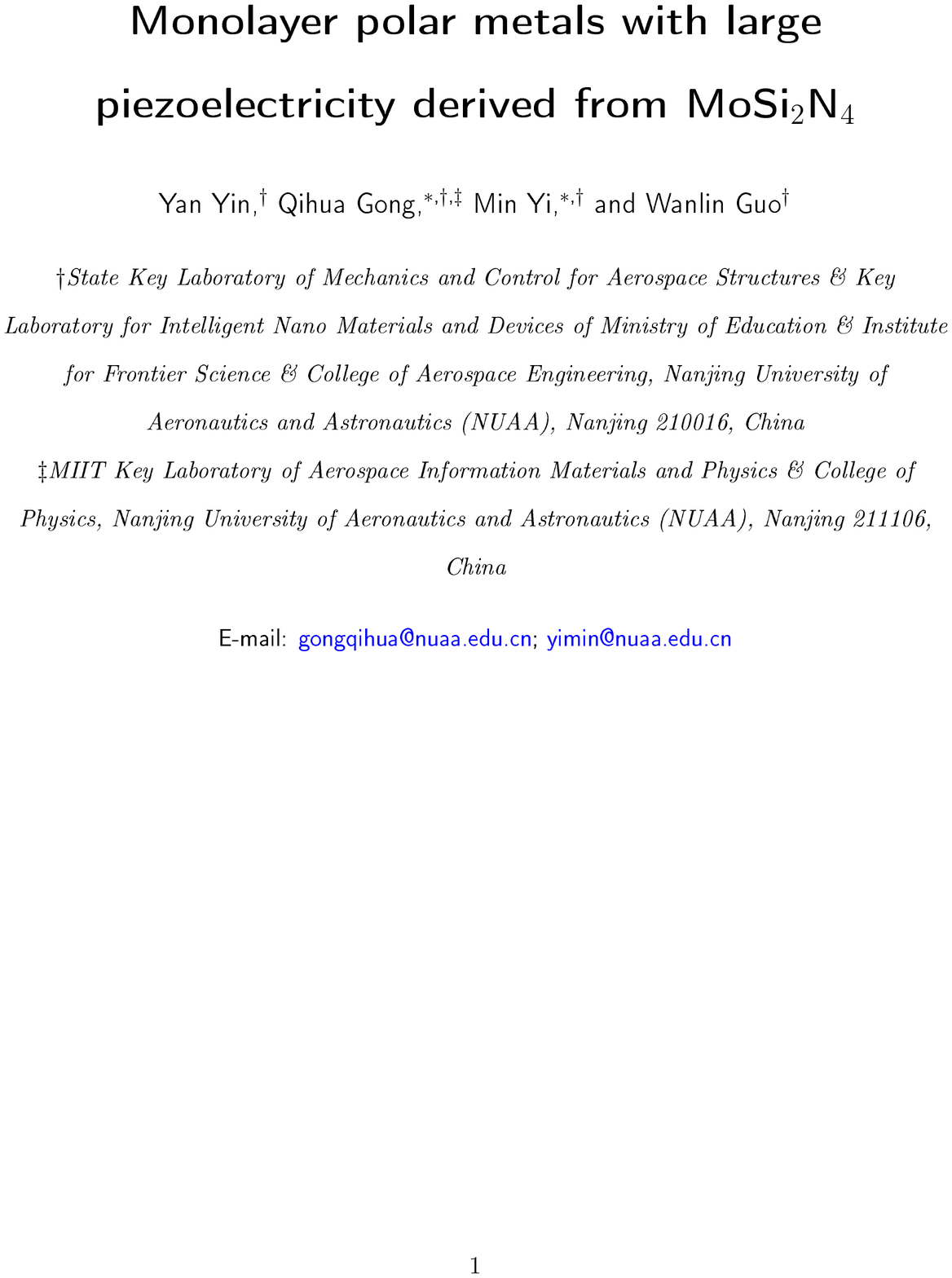}

\end{document}